\documentclass{ragtime}

\usepackage{amsthm}
\usepackage[mathscr]{euscript}
\usepackage{times}
\usepackage{txfonts}
\usepackage{color}
\usepackage{dirtytalk}
\usepackage[T1]{fontenc}
\usepackage[utf8]{inputenc}
\usepackage[english]{babel}
\DeclareMathOperator*{\argmax}{argmax} 

\usepackage{soul,ulem}

\definecolor{darkgreen}{rgb}{0.0, 0.75, 0.0}

\title[Extreme Mass Ratio Inspirals: accessing generic orbits]
      {Extreme Mass Ratio Inspirals:\\accessing generic orbits}

\author[O. Zelenka et al.
            ]
       {Ondřej Zelenka\ins{1,a}, Michal Straten\'{y}\ins{1}, Petr Kotla\v{r}\'{i}k\ins{1} and\splitauthors Georgios Lukes-Gerakopoulos\ins{1,}\ins{~2}\\ 
        \\
        \ins{1} \hspace{-0.5mm}Astronomical Institute,
       Czech Academy of Sciences,\splitins[2]
       \hspace{-0.5mm}Boční II 1401/1a, CZ-141 00 Prague, Czech Republic\\
       \hspace{-0.5mm}\ins{2} \hspace{-0.5mm}Department of Physics, University of Thessaly,
       GR-35100 Lamia, Greece\\
        \ins{a}\hspace{-0.5mm}\Email{zelenka@asu.cas.cz}} 

\providecommand{\dif}{\dif} 

\begin{document}

	\begin{abstract}

Extreme Mass Ratio Inspirals (EMRI) are a significant gravitational wave source and a target for the upcoming LISA mission. Computations of the fluxes of conserved quantities, such as energy and angular momentum, are a critical part of modeling EMRI waveforms and can be performed using a frequency domain formalism to solve the Teukolsky equation. This requires computations of large numbers of mode amplitudes, many of which are negligible. However, it is unknown ex ante which can be safely skipped. We present a machine learning approach to estimate the dominant radial mode number as a function of the remaining mode numbers and apply it to reduce the number of unnecessary computations in EMRI flux calculation.

	\end{abstract}

\begin{keywords}
	black hole~--~gravitational waves~--~Teukolsky equation~--~Extreme Mass Ratio Inspiral~--~machine learning
\end{keywords}

\section{Introduction}\label{sec:intro}

The LISA mission, planned to launch in 2035~\citep{Sala:2025edg}, is expected to observe gravitational waves from multiple types of sources, including Extreme Mass Ratio Inspirals (EMRI), which consist of a stellar-mass secondary compact object orbiting a supermassive black hole primary. They will spend up to months in the LISA sensitive band and be visible for thousands of orbits, allowing precision tests of the general theory of relativity~\citep{Babak:2017tow}.

Modeling an EMRI suffers from a fundamental issue: the disparity of the timescales involved. Given a primary mass $M$ and secondary mass $\mu$, the orbital times scale as $M$, while the gravitational radiation reaction drives the inspiral at times $\sim M^2/\mu$~\citep{Hinderer:2008dm}; at mass ratios of $\mu/M \lesssim 10^{-5}$, this disparity is sufficient to make numerical relativity simulations, which are very successful in modeling similar-mass binaries, practically infeasible.

Waveform modeling of EMRIs relies on the two-timescale approximation: averaging out phenomena on the orbital timescales and preserving long-term evolutions on the radiation reaction timescale~\citep{Hinderer:2008dm}. The radiation reaction effects are incorporated by treating the secondary as a test particle and computing fluxes of conserved quantities using black hole perturbation theory. This amounts to solving the Teukolsky equation and extracting the fluxes at the horizon and null infinity. Using this method, fast generation of EMRI waveforms has been achieved for eccentric equatorial orbits~\citep{Chua:2020stf,Chapman-Bird:2025xtd}.

The Teukolsky equation admits solutions in the frequency domain by separation into modes and allows for straightforward extraction of fluxes \citep{Hughes:2021exa}. However, including a generic geodesic orbit as a source produces a 4-dimensional mode expansion without much knowledge of which modes are relevant. This is especially true for the radial mode expansion for high eccentricity orbits, even producing multiple local maxima along a single axis. This means that choosing a suitable mode truncation procedure is somewhat tricky and in most cases, large numbers of irrelevant modes need to be computed to ensure accuracy of the method.

In the eccentric equatorial case, the mode expansion is only 3-dimensional and so is the parameter space. Extending to the generic case (inclined and equatorial inspirals) increases the dimensionality of both of these, which makes the computation of the fluxes on a grid computationally much more expensive~\citep{Chapman-Bird:2025xtd}.

In this paper, we attempt to mitigate this issue by introducing a method to map the space of modes of a generic EMRI orbit and only solve for the relevant modes. It consists of an estimator of the dominant radial mode number given the other mode numbers based on elementary machine learning techniques. For a given orbit, this would allow taking any combination of the other mode numbers and only calculating the fluxes of the corresponding dominant radial mode; if it is sub-threshold, the entire combination can be discarded.

The paper is organized as follows: in Sec.~\ref{sec:fluxes_bg} we describe the fluxes calculation procedure to an elementary degree in order to provide a background for the particular mode expansion. In Sec.~\ref{sec:nn} we present the composition and training procedure of a neural network and in Sec.~\ref{sec:estimator} we build the radial mode estimator. In Sec.~\ref{sec:conclusion} we summarize our findings and the next steps towards a more robust result.

\section{Teukolsky equation fluxes}\label{sec:fluxes_bg}

\subsection{Frequency domain fluxes}

This section presents a brief look into the frequency domain formalism to solve the Teukolsky equation used in this work \citep[see e.g.][]{Hughes:2021exa}. Let us list the most important elements.

The Teukolsky equation governs linear perturbations of a Kerr black hole via the Newman-Penrose curvature scalar $\psi_4$. Using Boyer-Lindquist coordinates, it decomposes into $l, m$ modes through the $s=-2$ spheroidal harmonics $S_{lm}$ as
\begin{equation}
    \psi_4 = \frac{1}{\left(r - ia\cos\theta\right)^4} \int_{-\infty}^\infty \dif\omega\sum_{m = -\infty}^{\infty}\sum_{l = |m|}^{\infty} R_{lm}\left(r,\omega\right)S_{lm}\left(\theta, a\omega\right)e^{i\left(m\phi - \omega t\right)} ~,
\end{equation}
leaving us to solve an ordinary differential equation for the radial functions $R_{lm}$.

The motion of the test particle is described by geodesics in the Kerr spacetime background. The time and azimuthal coordinates separate out trivially due to symmetry, and with reparametrization in terms of the Mino time $\lambda$ defined by $\dif\lambda = \dif\tau/\Sigma$ using the metric function $\Sigma = r^2 + a^2\cos^2\theta$ the radial and polar degrees of freedom also decouple into the form~\citep{Schmidt:2002qk}
\begin{equation}
    \left(\frac{\dif\theta}{\dif\lambda}\right)^2 = \Theta\left(\theta\right)~,\quad \left(\frac{\dif r}{\dif\lambda}\right)^2 = R\left(r\right) ~.
\end{equation}
Their evolution is, thus, described by periodic functions (and so is $\phi$) with frequencies $\Upsilon_\theta$, $\Upsilon_r$, and $\Upsilon_\phi$ with respect to Mino time and $\Omega_\theta$, $\Omega_r$, $\Omega_\phi$ with respect to coordinate time. They can be expanded in Fourier series
\begin{equation}
    \theta\left(\lambda\right) = \sum_{k=-\infty}^{\infty} \theta_k e^{-ik\Upsilon_\theta \lambda} ~,\quad r\left(\lambda\right) = \sum_{n=-\infty}^{\infty} r_n e^{-in\Upsilon_r \lambda} ~.
\end{equation}
The orbit is described using a variant of the usual orbital parameters: semilatus rectum $p$, eccentricity $e$, and inclination $I$~\citep{Schmidt:2002qk}. Frequently the parameter $x=\cos I$ is used instead.

After constructing the source term using these expressions, we arrive at the expressions
\begin{subequations}
\begin{align}
    \left(\frac{\dif E}{\dif t}\right)^\infty = \sum_{l,m,k,n} \left(\frac{\dif E}{\dif t}\right)_{lmkn}^\infty ~,\quad \left(\frac{\dif E}{\dif t}\right)_{lmkn}^\infty = \frac{\left|Z_{lmkn}^\infty\right|^2}{4\pi\omega_{mkn}^2} ~.
\end{align}
\end{subequations}
Similar expressions exist for the infinity fluxes of the azimuthal angular momentum and the Carter constant, and for the horizon fluxes of all three integrals of motion.

The amplitudes $Z_{lmkn}^{\infty}$ are determined by solving the radial Teukolsky equation and extracting their asymptotic values at the horizon and infinity, and the mode frequency is
\begin{equation}
    \omega_{mkn} = m\Omega_\phi + k\Omega_\theta + n\Omega_r ~.
\end{equation}

Precision calculations of the amplitudes can be achieved using community packages such as the \verb|Teukolsky| module of the Black Hole Perturbation Toolkit package for Mathematica~\citep{BHPToolkit, BHPToolkit:Teukolsky} or the \verb|pybhpt| package in C++ and Python~\citep{Nasipak:2025tby,Nasipak_pybhpt_2026}.

\begin{figure}[ht]
    \begin{center}
        \includegraphics[width=.85\linewidth]{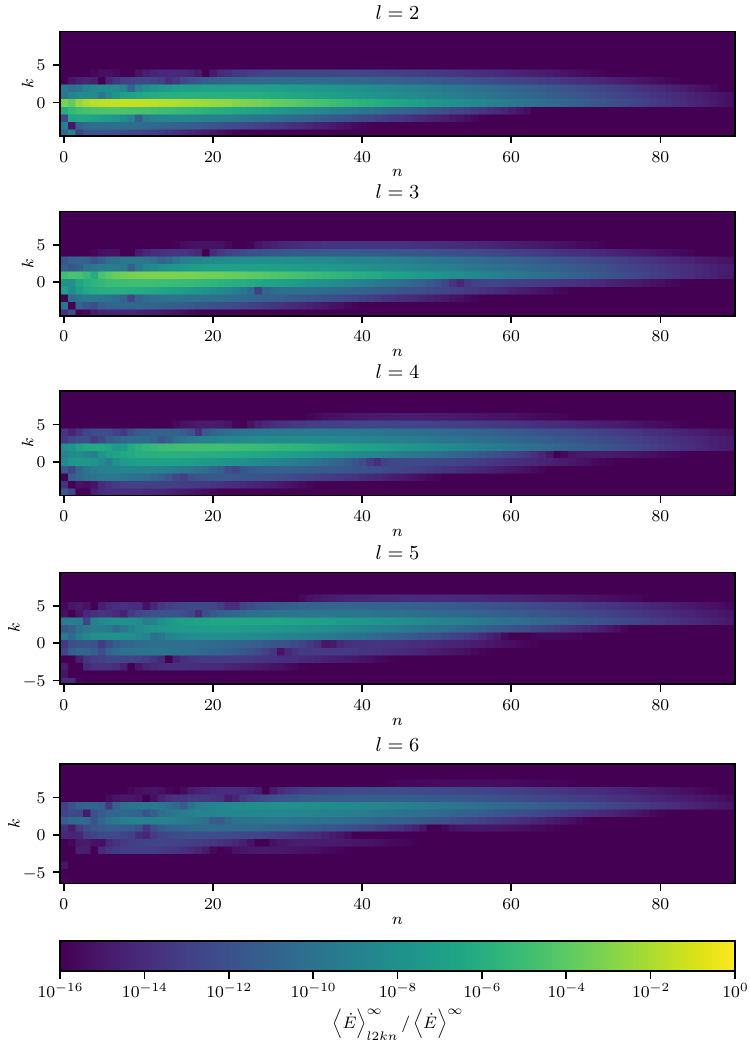}
        \caption{Example plot of energy contributions of various modes. The parameters of the corresponding orbit are $a=0.9$, $p=13.1982M = p_\mathrm{LSO}+10M$, $e=0.7$, $I=20^\circ$. For each $l$, all modes with $m=2$, $k \in \left[\min\left(-l,-4\right), 9\right]$, $n \leq 90$ are calculated and their respective relative contributions to energy fluxes to infinity are shown in an individual panel. Notably there are local dark spots which correspond to local minima with respect not only to the radial expansion in $n$ but also the polar expansion in $k$. The mode amplitudes and fluxes were calculated using \texttt{pybhpt} for illustration.
            \label{fig:modes_m2_pcolor}}
    \end{center}
\end{figure}

\begin{figure}[ht]
    \begin{center}
        \includegraphics[width=.85\linewidth]{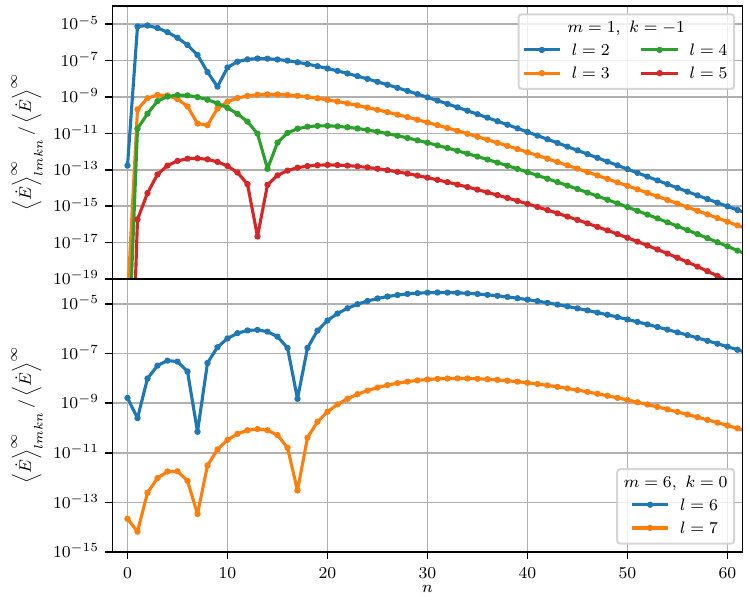}
        \caption{Example plot of energy contributions of various modes. The parameters of the corresponding orbit are $a=0.9$, $p=13.1982M = p_\mathrm{LSO}+10M$, $e=0.7$, $I=20^\circ$. In the top panel, modes $m=1$, $k=-1$ are shown as a function of $n$, with four curves for $l=2,3,4,5$ respectively. In the bottom panel, modes $m=6$, $k=0$ are shown as a function of $n$, with two curves for $l=6,7$ respectively. The curves feature multiple local maxima and minima, which are an issue for determining a suitable mode truncation for efficient flux calculation. The mode amplitudes and fluxes were calculated using \texttt{pybhpt} for illustration.
            \label{fig:modes_lines}}
    \end{center}
\end{figure}

As the expansion occurs along 4 parameters, including higher modes causes the computational cost to rise very quickly. Furthermore, the mode contributions to energy fluxes peak at various values of $n$ for different values of $l, m, k$ and orbital parameters; an example of this is shown in Fig.~\ref{fig:modes_m2_pcolor} and, in a more limited but clearer manner, Fig.~\ref{fig:modes_lines}. Truncating the mode sum with traditional methods thus often involves computations of large numbers of unnecessary modes.

\subsection{Datasets}

In this work, data on dominant radial $n$ modes are generated using the \verb|Teukolsky| module of the Black Hole Perturbation Toolkit~\citep{BHPToolkit,BHPToolkit:Teukolsky}, and verified against an independent numerical code \verb|pybhpt|~\citep{Nasipak:2025tby,Nasipak_pybhpt_2026}, and against the 4PN expansion of \citet{sago2015:Calculationradiation} within its domain of validity.

We choose the orbital parameters from a grid over
\begin{itemize}
    \item for all orbits $a=0.9M$,
    \item $e=0.1, 0.4, 0.7$,
    \item $I = 20^\circ, 50^\circ, 80^\circ$,
    \item $p - p_\mathrm{LSO}\left(a, e, x\right) = 10M, 32.5M, 55M, 77.5M, 100M$.
\end{itemize}
The values of $p$ are distributed at uniform distances from the last stable orbit (LSO), whose position is dependent on $a, e, x$~\citep{Stein:2019buj}. This grid forms a total of 45 orbits.

At each grid point, amplitudes and their corresponding contributions to the energy fluxes are calculated to a precision of $10^{-4}$, meaning that the sum of all neglected modes contributes at most $\mathcal{O}(10^{-4})$ to the total fluxes. We truncate the mode sum one index at a time, based on how the fluxes behave empirically in each index. For fixed $m$, $l$, $k$, the fluxes first oscillate in $n$, reaching a maximum at a single $n_\text{max}$, after which they decay roughly as a power law. We therefore start at $n = 0$ and stop once the last 20 modes are all decreasing and each contributes below the threshold. Next, for fixed $m$, $l$, we sum over $n$ for each $k$. For $m = \pm 2, \pm 1$, and $0$, the resulting fluxes as functions of $k$ typically have two maxima at $k_\text{max} < k^\prime_\text{max}$, with $k_\text{max} \approx -l - m$ and $k^\prime_\text{max} \approx l - m$. Outside the interval $[k_\text{max}, k^\prime_\text{max}]$, the fluxes decay rapidly, so we truncate the sum over $k$ in both directions once the contributions drop below the threshold. Between the two maxima, however, the contributions can stay roughly constant, so we compute all modes in this interval. For modes $m>2$, only the maximum $k^\prime_\text{max}$ remains, and for modes $m < -2$ only the one at $k_\text{max}$. We therefore start at this maximum and proceed in both directions until two consecutive $k$ modes, each summed over $n$, fall below the threshold. Finally, we stop increasing $l$ once the contribution of a given $l$, summed over all $k$ and $n$, falls below the threshold. Similarly, we stop increasing $m$ once the contribution of a given $m$, summed over all $l$, $k$ and $n$, falls below the threshold. We start with the most dominant modes, $m = \pm2$, $\pm 1$, and $0$.

A total of 8798 $l, m, k$ are thus calculated, corresponding to a mean of $\sim 195.5$ per orbit. For every $l, m, k$ combination a long segment of $n$ values starting at 0 is chosen and the corresponding modes are calculated. The label of each data point is then
\begin{equation}
    n_\mathrm{max}\left(a, p, e, x, l, m, k\right) = \argmax_{n} \left(\frac{\dif E}{\dif t}\right)_{lmkn}^\infty ~.
\end{equation}

All the data points are then shuffled and randomly split into:
\begin{itemize}
    \item 60\% training dataset,
    \item 20\% validation dataset,
    \item 20\% test dataset.
\end{itemize}

\section{Neural networks}\label{sec:nn}

An artificial neuron is the stacking of a linear operation with multiple inputs and a non-linear \textit{activation function} which is, in most cases, applied to scalar outputs. The coefficients of the linear operation are called \textit{weights}.

A simple neural network (NN) is built by organizing several neurons (with independent weights) into layers and stacking these so that the output of each feeds into the next as input (besides the final layer). Such a network can be formulated as a sequence of matrix multiplications alternating with the non-linear activation. With arbitrary numbers of layers and layer sizes, a neural network is a universal approximant: it can approximate any function to an arbitrary precision (with suitable choices for the error function), as shown by~\cite{Hornik1989,Cybenko1989}. At the same time, this is not always practically possible; this is known as the ``no free lunch'' theorem~\citep{Wolpert1996,Wolpert1997}.

To obtain suitable values for the network weights (summarily called the network state), one must optimize an error function over some sort of data. In the supervised learning approach, these consist of input/output pairs. In this work we have used the mean squared error loss function for model outputs $\mathbf{Y}$ and labels $\bar{\mathbf{Y}}$
\begin{equation}
    \mathrm{MSE}\left(\mathbf{Y}, \bar{\mathbf{Y}}\right) = \frac{1}{N}\sum_{i=1}^N \left(Y_i - \bar{Y}_i\right)^2 ~.
\end{equation}
At the same time, the optimization may lead to network states which fit the training data very well but fail to generalize to more data following the same distribution. This phenomenon is called \textit{overfitting} and is countered by verifying the network performance on a second independent \textit{test dataset}.

\begin{figure}[ht]
    \begin{center}
        \includegraphics[width=0.4\linewidth]{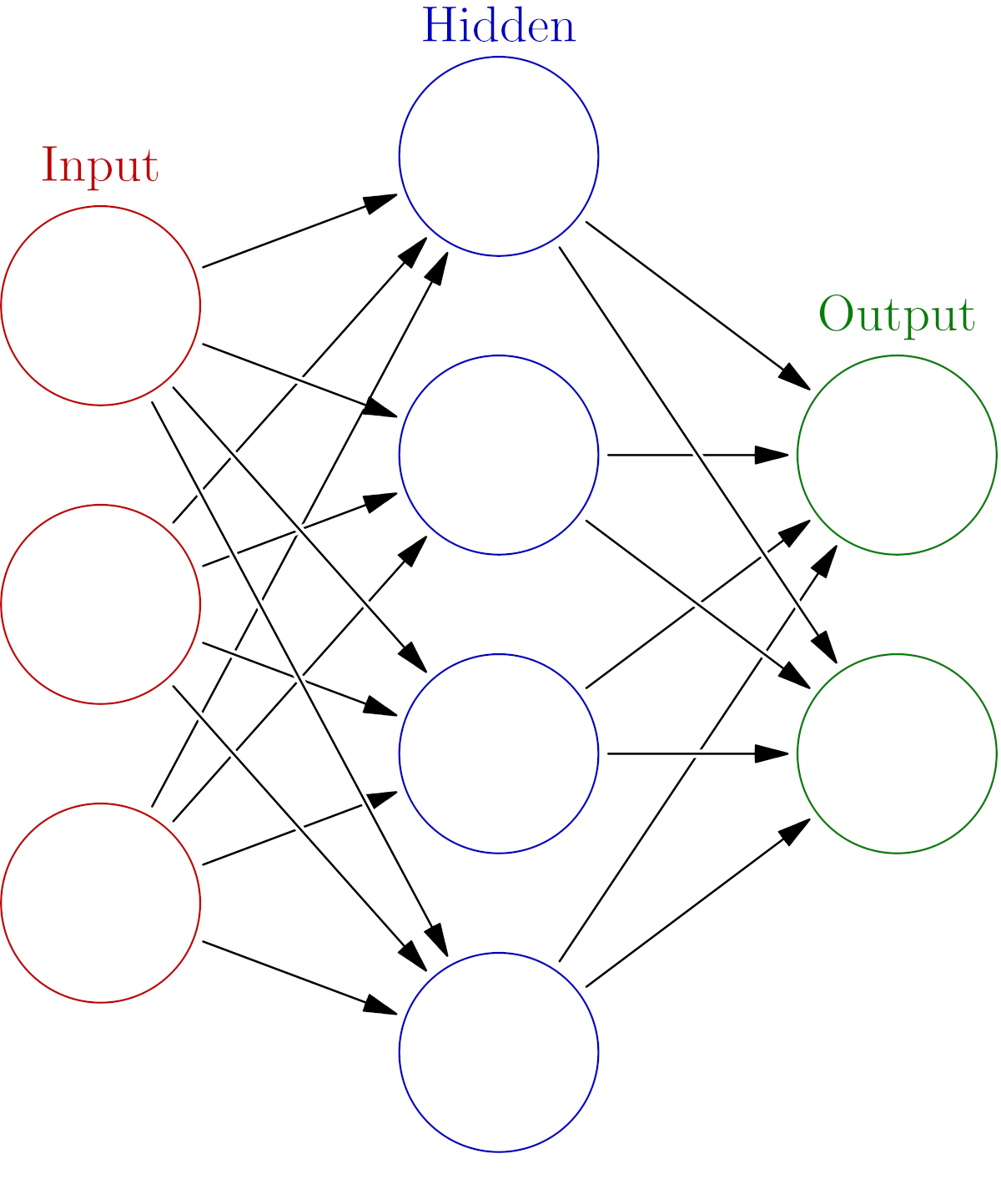}
        \caption{Example schematic of a neural network. Source: \href{https://commons.wikimedia.org/wiki/File:Colored_neural_network.svg}{Glosser.ca}, \href{https://creativecommons.org/licenses/by-sa/3.0}{CC BY-SA 3.0}, via Wikimedia Commons
            \label{fig:nn}}
    \end{center}
\end{figure}

In most cases, a gradient descent based approach is used which optimizes the error function by tuning the network weights iteratively. Many methods include momentum and/or a running mean of the gradient and/or momentum to improve convergence as the landscape of an error function over a dataset often has many local minima. We use the Adam optimizer~\citep{kingma2015adam} which includes an adaptive learning rate. Furthermore, to accelerate training, stochasticity is introduced by splitting the training dataset into batches. A single training epoch then consists of one optimizer step over each training batch.

With iterative optimizers overfitting often occurs at late learning stages. To mitigate this a third \textit{validation dataset} is used. The validation loss is calculated at every step of the optimization to monitor when the network begins to overfit and select an optimal network state.

\section{Dominant radial mode estimation}\label{sec:estimator}

In many cases the $l = 2, m = \pm 2$ modes dominate the gravitational radiation at infinity. The question arises naturally: is there a formula to find dominant modes in the full $\left(l, m, k, n\right)$ decomposition?

\begin{figure}[ht]
    \begin{center}
        \includegraphics[width=0.8\linewidth]{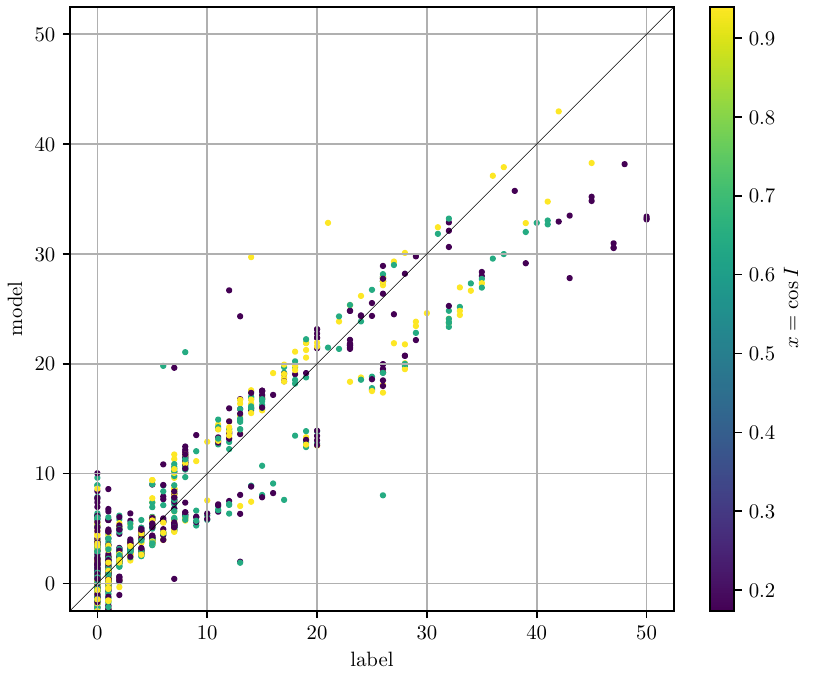}
        \caption{Performance of the simple linear model on the test dataset. Each point is drawn with the label (true $n_\mathrm{max}$ on the horizontal axis) and the model output on the vertical axis. Points are colored by their $x=\cos I$ and the diagonal is drawn to show perfect accuracy.
            \label{fig:model_00}}
    \end{center}
\end{figure}

Let us begin with a simple model where the $n_\mathrm{max}$ is a linear function of $l, m, k$, while the coefficients may depend on the orbital parameters. The model is thus
\begin{subequations}
\begin{align}
    n_\mathrm{max} &= \alpha l + \beta m + \gamma k ~,\\
    \alpha, \beta, \gamma &= \mathrm{NN}\left(a, p, e, x\right) ~.
\end{align}
\end{subequations}
The $\mathrm{NN}$ is a simple neural network consisting of three linear layers of intermediate size 10 and using the Rectified Linear unit activation function
\begin{equation}
    \mathrm{ReLU}\left(x\right) = \begin{cases}x, & x \geq 0 \\ 0, & x < 0\end{cases} ~.
\end{equation}

The network is then trained using the Adam optimizer with a base learning rate of $10^{-2}$ and a batch size of 64. Multiple setups were attempted to optimize both the model and the training procedure. Both the training and the validation loss converge to a value of $\mathrm{MSE}\approx 10$, overfitting is therefore not an issue.

The performance of the model is shown in Fig.~\ref{fig:model_00} by directly comparing outputs with labels. While values are clustered along the main diagonal corresponding to perfect performance, a large fraction of points deviate by more than 5. The model thus fails to capture the complexity of the behavior of the mode expansion, and we conclude that the target function is non-linear with respect to the $l, m, k$ arguments.

\begin{figure}[ht]
    \begin{center}
        \includegraphics[width=0.8\linewidth]{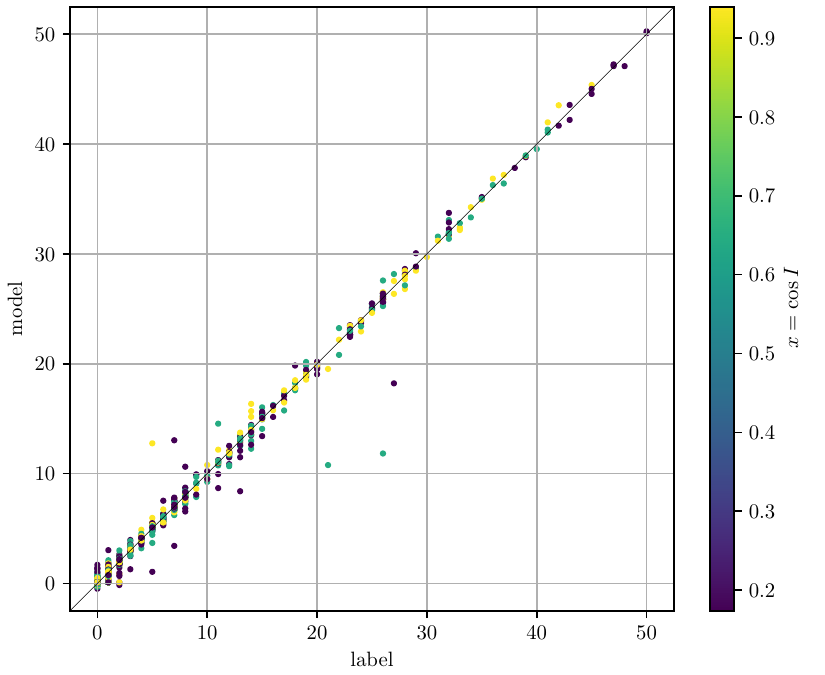}
        \caption{Performance of the complex non-linear model on the test dataset. Each point is drawn with the label (true $n_\mathrm{max}$ on the horizontal axis) and the model output on the vertical axis. Points are colored by their $x=\cos I$ and the diagonal is drawn to show perfect accuracy.
            \label{fig:model_18}}
    \end{center}
\end{figure}

Let us attempt to train a more complex network, non-linear with respect to all seven arguments
\begin{equation}
    n_\mathrm{max} = \mathrm{NN}\left(a, p, e, x, l, m, k\right) ~.
\end{equation}
In this case we use a network consisting of 4 layers with intermediate input/output sizes of 20, 50, and 20. Between the layers the Gaussian Error linear unit activation is used
\begin{equation}
    \mathrm{GELU}\left(x\right) = x\cdot \Phi\left(x\right) ~,
\end{equation}
where $\Phi\left(x\right)$ is the cumulative distribution function of the Gaussian distribution. It has the same asymptotic behavior as $\mathrm{ReLU}$ but is smooth over $\mathbb{R}$ which improves training stability. This network is trained over the same training dataset with a batch size of 16 and a learning rate of $5\cdot 10^{-3}$.

In multiple training runs with independent initial weights, the training proceeds smoothly and reaches validation loss values of $\approx 0.4$, which are lower than in the simple linear case by more than an order of magnitude. One of the training runs produces a higher minimal validation loss than the lowest by a negligible amount while reducing the number of outliers, which we choose as the final state.

Fig.~\ref{fig:model_18} shows the performance of this model on the test dataset. The vast majority of points are clustered very close to the diagonal and only a few visibly stray far from it (out of a total of 1760 points). Studying the outliers reveals no correlation with orbital parameters. However, there may be a connection to the mode numbers, as high $l$ and $m$ values seem to correlate with high errors.

When the network output is rounded to the nearest integer, for 88.6\% of the points the model outputs match the label exactly, and with a $\pm 1$ tolerance 98.5\% of the points produce the correct output. Increasing the tolerance to 3, only 9 out of 1760 points (0.5\%) deviate by 4 or more.

	

\section{Conclusions}\label{sec:conclusion}

We have presented a neural network trained to estimate the dominant radial mode in an Extreme Mass Ratio Inspiral (EMRI) snapshot orbit to reduce the number of required computations. While the model seems to be accurate on the data at hand, further development is required before it can be used in production runs for EMRI flux datasets. Most importantly, all the data were generated using a primary black hole spinning at $a=0.9M$, while a full-fledged model would have to cover a wide range including negative values.

Furthermore the same orbits are included in the training, validation and test datasets. While these are separate data points since the same orbit would appear with different $\left(l, m, k\right)$ combinations in different datasets, independent verification by additional previously unseen orbits is required.

Similar frequency domain approaches are also used to calculate higher order terms in the postadiabatic expansion, such as the secondary spin contributions. The machine learning approach presented in this work can be adapted to these cases as well. However, retraining the network will be necessary, as the mode structure of the spin contributions and others is different.

\ack

O. Z. thanks the Czech Academy of Sciences for funding through the PPLZ fellowship. O.Z., M. S. and P. K. are supported by the LUC25003 InterCOST program of the Ministry of Education, Youth and Sports of the Czech Republic (MŠMT). O. Z. also acknowledges support from the programme Strategy AV21, project ``AI: Artificial Intelligence for Science and Humanity''. This work makes use of the Black Hole Perturbation Toolkit.


\def\prc{Phys. Rev. C}
\def\pre{Phys. Rev. E}
\def\prd{Phys. Rev. D}
\def\jcap{Journal of Cosmology and Astroparticle Physics}
\def\apss{Astrophysics and Space Science}
\def\mnras{Monthly Notices of the Royal Astronomical Society}
\def\apj{The Astrophysical Journal}
\def\aap{Astronomy and Astrophysics}
\def\actaa{Acta Astronomica}
\def\pasj{Publications of the Astronomical Society of Japan}
\def\apjl{Astrophysical Journal Letters}
\def\pasa{Publications Astronomical Society of Australia}
\def\nat{Nature}
\def\physrep{Physics Reports}
\def\araa{Annual Review of Astronomy and Astrophysics}
\def\apjs{The Astrophysical Journal Supplement}
\def\aapr{The Astronomy and Astrophysics Review}
\def\procspie{Proceedings of the SPIE}

\def\prc{Phys. Rev. C}
\def\pre{Phys. Rev. E}
\def\prd{Phys. Rev. D}
\def\jcap{Journal of Cosmology and Astroparticle Physics}
\def\apss{Astrophysics and Space Science}
\def\mnras{Mon. Not. R. Astron Soc.}
\def\apj{The Astrophysical Journal}
\def\aap{Astron. Astrophys.}
\def\actaa{Acta Astronomica}
\def\pasj{Publications of the Astronomical Society of Japan}
\def\apjl{Astrophysical Journal Letters}
\def\pasa{Publications Astronomical Society of Australia}
\def\nat{Nature}
\def\physrep{Phys. Rep.}
\def\araa{Annu. Rev. Astron. Astrophys.}
\def\apjs{The Astrophysical Journal Supplement}
\def\aapr{The Astronomy and Astrophysics Review}

\def\mdash{---}
	\bibliography{references}
	
\end{document}